\documentclass[twocolumn,pre,showpacs,amsmath,amssymb,floatfix]{revtex4-1}

\usepackage{graphicx}
\usepackage{dcolumn}
\usepackage{bm}
\usepackage{amsmath,amssymb}

\begin{document}

\title{Space-Time-Resolved Capillary Wave Turbulence}

\author{Michael Berhanu}
\author{Eric Falcon}
\affiliation{Mati\`ere et Syst\`emes Complexes (MSC), Universit\'e Paris Diderot, CNRS (UMR 7057), 75013 Paris, France} 

\date{\today}

\begin{abstract}
 We report experiments on the full space and time resolved statistics of capillary wave turbulence at the air-water interface. The three-dimensional shape of the free interface is measured as a function of time by using the optical method of \textit{Diffusing Light Photography} associated with a fast camera. Linear and nonlinear dispersion relations are extracted from the spatio-temporal power spectrum of wave amplitude. When wave turbulence regime is reached, we observe power-law spectra both in frequency and in wave number, whose exponents are found in agreement with the predictions of capillary wave turbulence theory. Finally, the temporal dynamics of the spatial energy spectrum highlight the occurrence of stochastic bursts transferring wave energy through the spatial scales.
\end{abstract}

                             
\maketitle

\section{Introduction}
Wave turbulence concerns the study of the statistical properties of a set of numerous waves in nonlinear interaction. For strong enough interactions, a turbulent cascade transfers wave energy from an injection scale towards a dissipation scale. This phenomenon is analytically described in a weakly nonlinear regime by the weak turbulence theory. Analytical expressions of the wave spectrum as a scale power-law is then derived in out-of-equilibrium situations in nearly all fields of physics involving waves (for reviews see \cite{Zakharovbook,Nazarenkobook}). Several recent studies have tested the relevance of wave turbulence theory in well controlled laboratory experiments notably for bending waves in elastic plates~\cite{Mordant2008} and for hydrodynamic waves on the surface of a fluid in the gravity and capillary regimes (for a  review see \cite{Falcon2010}). Most \textit{in situ} or laboratory measurements on wave turbulence involve time signals at a fixed location and show partial agreement with the theory. Spatio-temporal measurements of the turbulent wave amplitude are thus needed to investigate basic mechanisms of wave turbulence. For gravity wave turbulence, laboratory experiments involve measurements resolved in time and restricted to 1D space~\cite{Denissenko2007}, or simultaneously resolved in time and in 2D space~\cite{Herbert2010,Cobelli2011}. For the capillary regime, most of previous laboratory works have tried to isolate capillary wave turbulence from the gravity wave regime by using a parametric forcing~\cite{Putterman1996,Putterman1997,Henry2000,Brazhnikov2002,Snouck2009,Xia2010}, by operating under microgravity~\cite{FalconFalcon2009} or by studying waves at the interface between two fluids of same density~\cite{During2009}. 

To our knowledge, there is no experiment studying capillary wave turbulence with a wave amplitude measurement simultaneously resolved in time and 2D space. As a consequence, linear and non linear dispersion relations have not been measured for capillary waves in a turbulent regime. Accurate characterization of capillary wave turbulence is of primmest interest at fundamental level. According to weak turbulence theory, energy transfer through the scales occurs by three-wave interactions, instead of four-wave interactions for gravity wave turbulence. Therefore, capillary and gravity wave turbulence differ fundamentally. Moreover, the validity domain of wave turbulence theory for capillary waves in experiments remains an open question, due to restrictive hypotheses of the theory (infinite isotropic and homogeneous system, weak non-linearity ...). In particular for scales smaller than or equal to the capillary length, viscous damping of waves is not negligible~\cite{Deike}, whereas theory relies on an Hamiltonian structure of the waves field. Moreover in oceanography, capillary waves dynamics could control heat and gaseous exchanges between ocean and atmosphere and contribute to the overall dissipation of gravity waves.

In this article, we report full space and time resolved experiments on the statistics of capillary wave turbulence generated by gravity waves, a forcing close to the oceanographic case. The free-surface elevation is directly measured by a sensitive optical method known as the \textit{Diffusing Light Photography} (DLP)~\cite{Putterman1996,Putterman1997}. This technique has a better spatial resolution and a higher sensitivity than the \textit{Fourier Transform Profilometry}  (an optical technique generally used to study wave turbulence at the gravity wave scales \cite{Herbert2010,Cobelli2011}), and is not limited to small wave steepness, in contrast to other optical methods measuring the local wave slope~\cite{Zhang1994,Moisy2009}. The DLP method was introduced more than 15 years ago to study capillary wave turbulence excited parametrically~\cite{Putterman1996}, and was able to get a 2D spatial measurement of the free surface amplitude but only at a given time (a CCD sensor recording only a photograph of the free surface). In our work, by associating the DLP technique with a fast camera nowadays available, we present the 2D spatial and temporal statistics of capillary wave turbulence excited by gravity waves. Note that recently, dynamics of highly non linear Faraday waves were studied using the DLP technique with a fast camera, but wave turbulence regimes have not been investigated~\cite{Xia2012}.  

\section{Experimental Setup}      
\begin{figure}[t!]
 \begin{center}
\includegraphics[width=8.6cm]{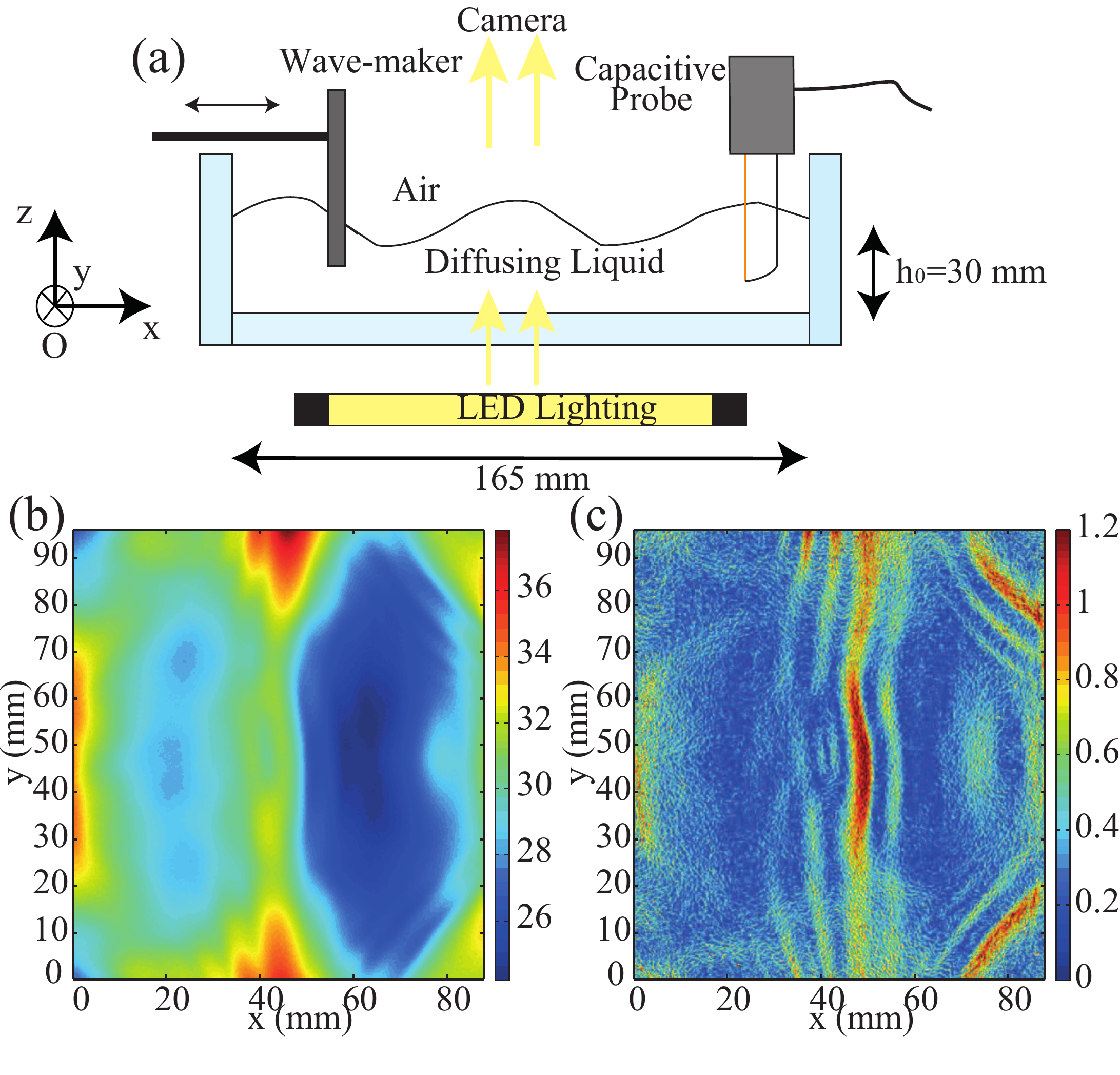}
 \caption{(color online) (a) Experimental setup (see text). (b) Snapshot of the wave field $h(x,y)$. Random forcing between 4 and $6\,$Hz. $\sigma_h=3.4$\,mm and $\sigma_s=0.30$ (see text). Colorscale is in mm. Wave maker is parallel to the $y$-axis and located at $x=-12$\,mm. (c) Snapshot of the spatial gradient of wave elevation $||\mathbf{\nabla}h(x,y)||$. Same parameters.}
    \label{schemab}
       \end{center}
 \end{figure}

The experimental setup is displayed in Fig.~\ref{schemab}(a). A Plexiglass tank ($165\times 165$\,mm$^2$) is filled with a diffusing liquid ($1$\,L of distilled water with $8$\,mL of Intralipid 20\%) up to height  $h_0=30\,$mm. Intralipid 20~\% (Fresenius Kabi \texttrademark)  is a commercial lipidic emulsion of microspheres, whose aqueous solutions are used as model diffusing media with characterized optical properties~\cite{Staveren}. Low dilution does not significantly modify the fluid viscosity and density from the pure water values ($\nu=10^{-6}$\,m/s$^{2}$ and $\rho=1000\,$kg/m$^{3}$, respectively). The value of surface tension $\gamma$ is obtained from the spatio-temporal measurements and is found to be $\gamma=60$ mN/m (see below). The transition between gravity and capillary waves is expected to occur for a critical wave number $k_c=\sqrt{\rho\,g/\gamma}$ which corresponds to a critical frequency $f_c \approx 14$\,Hz. Surface waves are generated by the horizontal motion of a rectangular paddle ($130$\,mm in width and $13 $\,mm in immersed depth) driven by an electromagnetic shaker (LDS V406) subjected to a random forcing (in phase and amplitude) band-pass filtered in frequency between $4$ and $6$\,Hz. By enhancing the initial mixing of waves, this type of forcing is known to produce cascades of gravito-capillary wave turbulence in laboratory experiments~\cite{Falcon2007,Herbert2010,Cobelli2011}. A LED device Phlox ($100 \times 100$\,mm$^2$) ensures homogeneous lighting below the transparent tank. A fast camera (PCO EDGE), one meter above, is focused on the liquid free-surface and records with $1024 \times 1120$ pixels and a $200\,$Hz frame rate on an observation area $\mathcal{S}$ of $89 \times 96$\,mm$^2$. Complementary measurements are also performed using a faster camera (Phantom V9) with a $1\,$kHz frame rate, but with lower sensitivity, and a limited number of images. For comparison, the wave amplitude at some location is also recorded by a capacitive wave gauge lying outside the camera field. 

Typical 2D spatio-temporal wave elevation is measured as follows. If the free mean path of light within the liquid is smaller than the depth of the fluid and larger than the wave amplitude, light going through the diffusing liquid is no more propagating like a ray, due to the scattering on micrometric particles. As a result, focusing of rays producing caustics should disappear~\cite{Putterman1997}. If the free mean path is also larger than the wave amplitude, the light amplitude $I(x,y)$ recorded by the camera on the free surface is controlled by the local depth of fluid $h(x,y)$~\cite{Putterman1997}. After static calibration on a flat surface, we find in agreement with Xia et al.~\cite{Xia2012}, an exponential decay of light amplitude with the local elevation 
\begin{equation}
 I(x,y)=I_0(x,y)\,\mathrm{exp}\left( - h(x,y)/{\mathcal{L}^*} \right)\,,
 \label{intens}
 \end{equation}
 $I_0$ being the light amplitude for $h=h_0$, and $\mathcal{L}^* \simeq 25\,$mm a typical distance of order of the free mean path of light within the liquid. If $\mathcal{L}^*$ is not too small compared to the fluid depth $h_0=30\,$mm, the multiple diffusion regime is not reached which guarantees the locality of the measurement and implies an exponential decay of light through the fluid~\cite{Ishimaru}.
The reconstruction of the free-surface topography $h(x,y,t)$ is then performed at each time step using Eq.\ \eqref{intens}. Note that images are spatially filtered with a Gaussian kernel of $5$ pixels (due to the camera photon shot noise) leading to a spatial resolution of roughly $0.5\,$mm. An example of free-surface reconstruction is shown in Fig.~\ref{schemab}(b). Gravity waves mainly propagate in the direction of wave maker vibrations. The corresponding spatial gradient of wave elevation $||\mathbf{\nabla}h||$, or wave steepness, is also shown in Fig.~\ref{schemab}(c). Movies of $h(x,y,t)$ and $||\mathbf{\nabla}h|| (x,y,t)$ are also shown in Ref.~\cite{Note1}. Although the wave field is found to be inhomogeneous and anisotropic, capillary waves will be found in a wave turbulence regime (see below). The full space and time resolved power spectrum of wave elevation $ S_h (\omega,k)$ is computed from the set of free-surface images $h(x,y,t)$, by performing successively a two-dimensional Fourier transform in space, and a Fourier transform in time, then integrating over the different directions of $\mathbf{k}$. This operation is performed on $8192$ (resp. $4196$) images corresponding to a duration of $41$\,s (resp. $4.2$\,s) for the 200 fps (resp. 1000 fps) camera. Two important parameters are the typical rms wave amplitude $$\sigma_h\equiv \left\langle \sqrt{\frac{1}{\mathcal{S}} \int_{\mathcal{S}} h^2(x,y,t) \mathrm{d}x \mathrm{d}y -\left( \frac{1}{\mathcal{S}} \int_{\mathcal{S}} h(x,y,t) \mathrm{d}x \mathrm{d}y \right)^2} \right\rangle $$ \\ and the typical rms wave steepness  $$\sigma_s \equiv \left\langle \sqrt{ \frac{1}{\mathcal{S}} \int_{\mathcal{S}} {||\mathbf{\nabla}h(x,y,t)||}^2 \mathrm{d}x \mathrm{d}y -\left( \frac{1}{\mathcal{S}} \int_{\mathcal{S}} {||\mathbf{\nabla}h||} \mathrm{d}x \mathrm{d}y \right)^2} \right\rangle, $$ where $\left\langle \cdot \right\rangle $ denotes a temporal averaging and $\int_{\mathcal{S}}$ a spatial integration on the surface $\mathcal{S}$.

\section{Linear and nonlinear dispersion relations}

\begin{figure}[t!]
 \begin{center}
 \includegraphics[width=8.6cm]{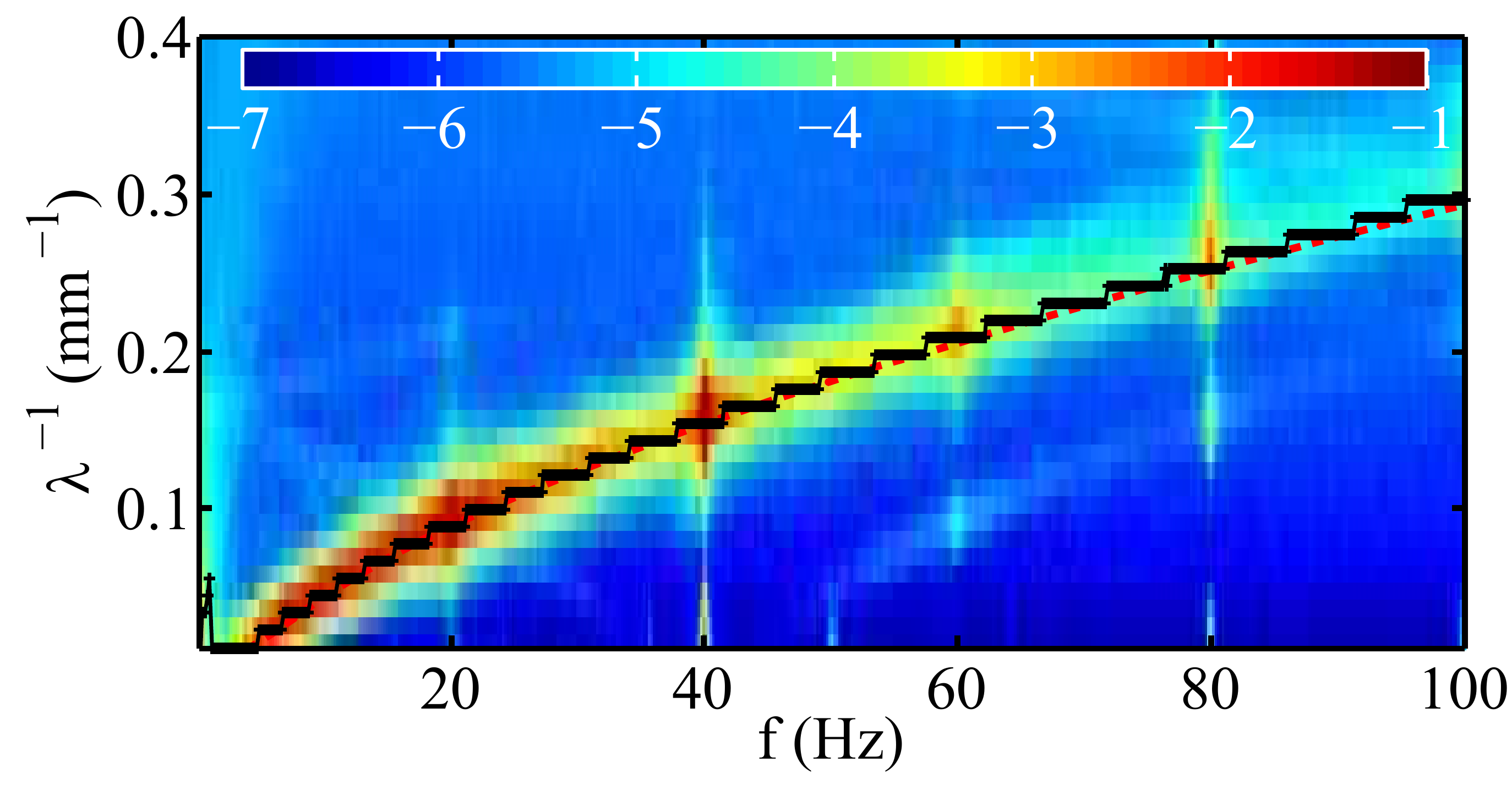} 
        \caption{(color online) Spatio-temporal spectrum of wave elevation $S_h (\omega,k)$ in sinusoidal forcing with an excitation frequency of $40$\,Hz. Amplitude $\sigma_h =0.22\,$mm and $\sigma_s=0.16$.  ($+$) Experimental dispersion relation extracted from the maxima of $S_h (\omega,k)$. Red (dark gray) dashed line: linear theoretical dispersion relation. Colorscale corresponds to $\log_{10}(S_h (\omega,k))$. Camera frame rate: $200$\,Hz.}
    \label{figsup}
       \end{center}
 \end{figure}

Let us first focus on the linear and nonlinear dispersion relation of capillary waves. For a random or a purely sinusoidal forcing at weak amplitude, the wave field should follow the linear gravity-capillary dispersion relation. In order to test this assumption, spatio-temporal power spectrum of wave elevation $S_h (\omega,k)$ is displayed in Fig.~\ref{figsup} for a sinusoidal forcing with an excitation frequency of $40$\,Hz. The maxima (black crosses) of the spectrum $S_h (\omega,k)$  give the experimental dispersion relation corresponding to the localization of wave energy in the $(\omega\equiv 2\pi f, k\equiv 2\pi/\lambda)$ space. $\lambda$ is the wavelength and $f$ is the frequency of waves. The experimental dispersion relation can be accurately fitted by the linear dispersion relation where the surface tension $\gamma$ is the only free parameter
\begin{equation}
\omega^2=\left[g\,k + (\gamma / \rho) \, k^3 \right]\tanh (k\,h_0)\, ,
\label{RDL}
\end{equation} 
$g=9.81$\,m/s$^2$ being the gravity acceleration and $\rho=1000$\,kg/m$^3$ the fluid density. We find $\gamma_{fit}=59.6\,$mN/m and then we consider $\gamma$ to be equal to $60$\,mN/m. Note that such a linear dispersion relation is observable, simultaneously at all scales, for strong enough forcing amplitude where nonlinearity spreads energy continuously in the $(\omega,k)$ space. Moreover, harmonic and subharmonic peaks are still visible (e.g. at $f=20$ Hz due to cross-wave instability~\cite{Barnard1972}), and the wave turbulence regime is thus not reached. 

\begin{figure}[h!]
 \begin{center}
 \includegraphics[width=8.6cm]{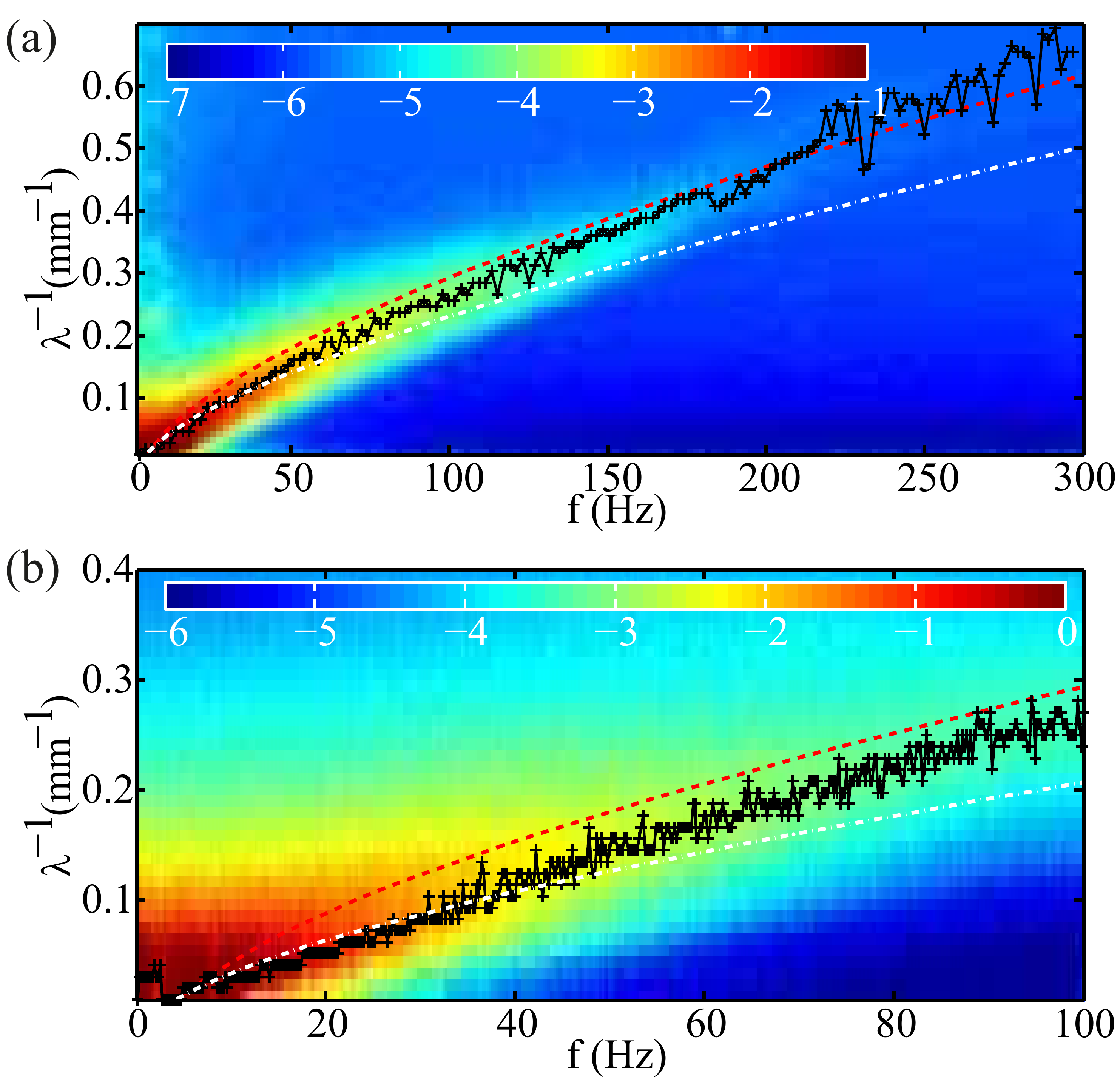} 
        \caption{(color online) (a) Spatio-temporal spectrum of wave elevation $S_h (\omega,k)$ for two different random forcing amplitudes: (a) $\sigma_h =2.7\,$mm and $\sigma_s=0.26$. (b) $\sigma_h =3.6\,$mm and $\sigma_s=0.34$. ($+$) Experimental dispersion relation extracted from the maxima of $S_h (\omega,k)$. Red (dark gray) dashed line: linear theoretical dispersion relation. White dot-dashed line: theoretical dispersion relation with a nonlinear shift (see text). Colorscales correspond to $\log_{10}(S_h (\omega,k))$. Camera frame rates: (a) $1$\,kHz and (b) $200$\,Hz.}
    \label{reladispMES6}
       \end{center}
 \end{figure}

At higher amplitudes of sinusoidal or random forcing, a departure from the linear dispersion relation is observed in particular in wave turbulence regimes as depicted in Fig.~\ref{reladispMES6}(a) for a moderate amplitude and in Fig.~\ref{reladispMES6}(b) for a stronger amplitude. The curve of the maxima of the spectrum (black crosses) differs significantly from the linear dispersion relation (red or dark gray dashed line). These discrepancy is found to increase with the forcing amplitude, and thus suggests a possible nonlinear shift of the dispersion relation. Indeed, dispersion relation of sinusoidal waves of high amplitude are theoretically known to experience a nonlinear shift in the gravity (Stokes waves)~\cite{Whitham} and capillary regimes~\cite{Crapper}. Although such a nonlinear shift is not predicted for nonlinear interacting waves with a continuous spectrum, an estimated nonlinear dispersion relation 
\begin{equation*}
\omega^2=\left(g k\left[1+(a k)^2\right]+\frac{\gamma}{\rho}k^3\left[1+\left(\frac{a k}{4}\right)^2\right]^{-1/4} \right)\tanh (k h_0)
\end{equation*}
 is also plotted in Fig.~\ref{reladispMES6} (white dot-dashed line), $a$ being the wave amplitude assumed to be equal to $\sigma_h$. The first term on the right-hand side of the equation (nonlinear gravity wave) dominates at low wave numbers and leads to a theoretical nonlinear dispersion relation below the linear one in the $(\omega, k)$ space. Experimental results follow a similar trend and suggest that nonlinear effects (appearance of Stokes waves) are the cause of the observed shift of the dispersion relation. Note that bound waves are not observed here contrary to larger size experiments studying the gravity wave turbulence regime~\cite{Herbert2010}. 
 
\begin{figure}[t!]
 \begin{center}
\includegraphics[width=8.6cm]{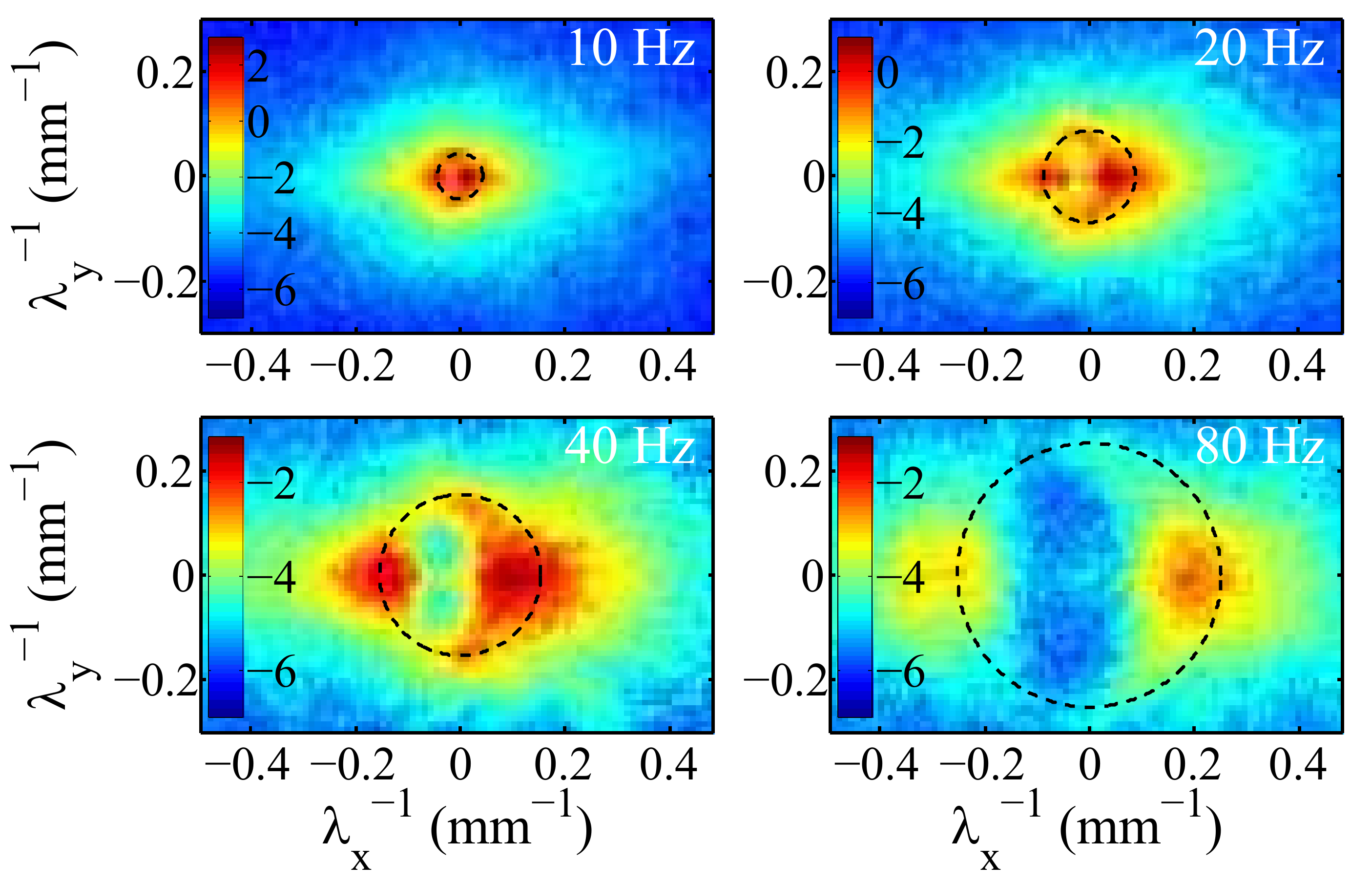} 
        \caption{(color online) Spectrum $S_h (\omega, {\lambda_x}^{-1}, {\lambda_y}^{-1})$ for fixed $f=10$, 20, 40 and 80 Hz ($log_{10}$ colorscale) showing anisotropy of wave field at high forcing amplitude ($\sigma_h =3.6\,$mm and $\sigma_s=0.34$). Dashed black circle: Linear dispersion relation of Eq.\ \eqref{RDL}. Wave amplitudes are larger in the direction of forcing ($x$-axis) revealing anisotropy in all frequency range.}
    \label{anisoMES6}
       \end{center}
 \end{figure}

Directional properties of the wave field are now investigated by plotting $S_h (\omega,k_x, k_y)$ as a function of  ${\lambda_x}^{-1}$ and ${\lambda_y}^{-1}$ for fixed $f=\omega /(2\pi)$ as in Fig.~\ref{anisoMES6}. Because the direction of the wave forcing is along the $x$-axis, we observe a strong anisotropy of the wave field. This anisotropy is conserved regardless of the frequency scales inside the capillary waves range (see at $f=20$, $40$ or $80$ Hz). Capillary waves appear preferentially along the same direction as the long waves from which they originate by nonlinear interactions. 
Due to the geometry of our experimental device and to the viscous dissipation, the wave field appears also inhomogeneous. The spatial inhomogeneity of the wave field can be estimated by taking the square root of the temporal average of $h^2 (x,y,t)$ and then computing the spatial standard deviation. We found in our experiments a spatial inhomogeneity of order of $24\%$, without significant changes with the wave amplitude.
 
  \begin{figure}[t!]
 \begin{center}
 \includegraphics[width=8.6cm]{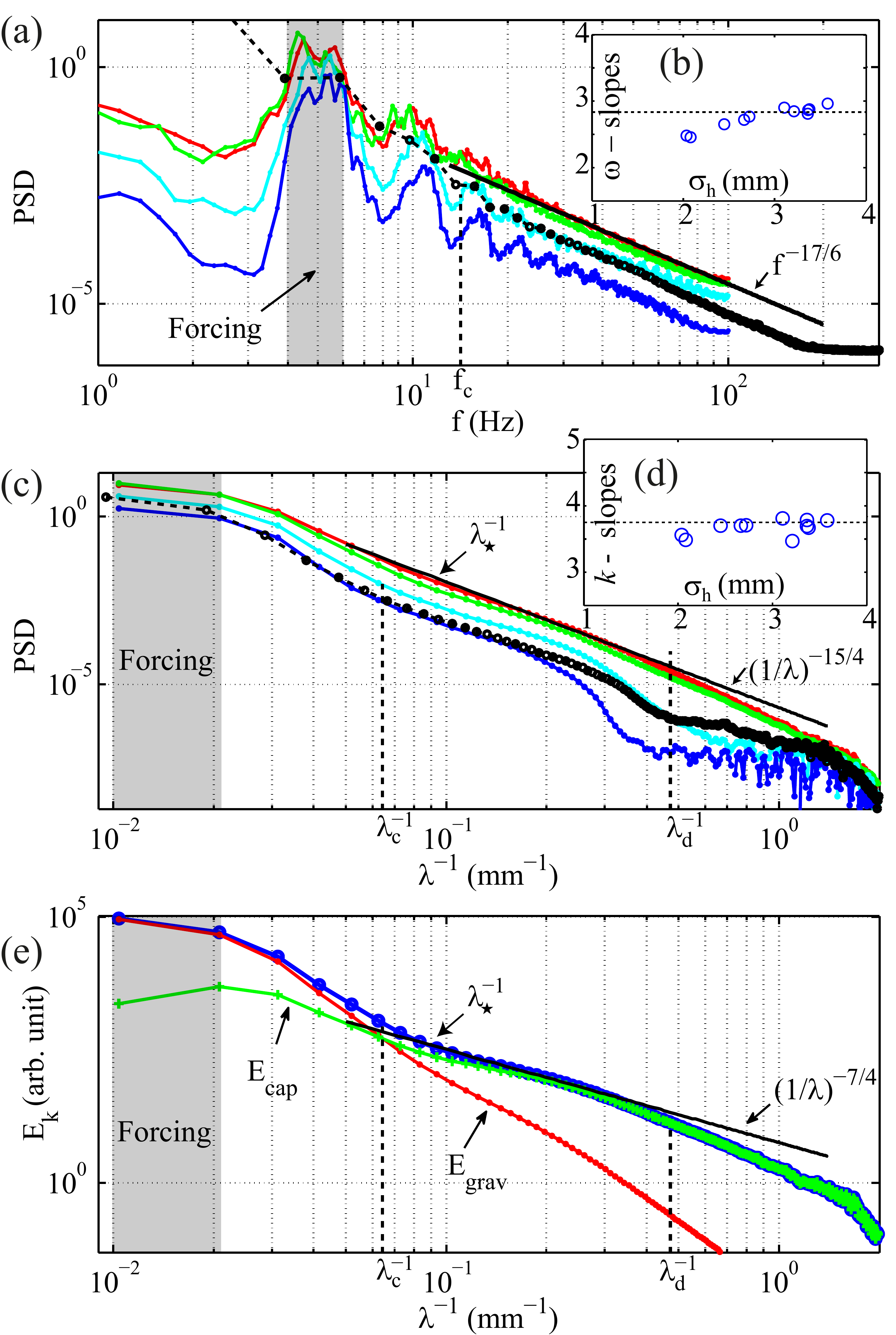} 
        \caption{(color online) (a) Temporal power spectra $S_h (\omega)$ for different forcing amplitudes. From bottom to top: $\sigma_h = $1.3, 2.7, 2.1, 3.4 and 3.6 mm, and $\sigma_s=0.15$, 0.26, 0.19, 0.3, and 0.34. Solid black line is the capillary prediction $f^{-17/6}$. The second measurement is recorded with a 1 kHz frame rate, the other ones at 200 Hz. (b) Inset: $S_h (\omega)$-exponents vs. $\sigma_h$ (fits from $20 \leq f \leq 100\,$Hz). Dashed line shows the theoretical value $-17/6$.  (c) Spatial power spectra $S_h (k)$ for the same measurements. Solid black line is the capillary prediction $k^{-15/4}$. (d) Inset: $S_h (k)$-exponents vs. $\sigma_h$ (fits from $0.094 \leq \lambda^{-1} \leq 0.30$ mm$^{-1}$). Dashed line shows the theoretical value $-15/4$. $\lambda_d$ is a characteristic scale of dissipation. $\lambda_c$ is the gravity-capillary crossover and $\lambda_{\star}^{-1}=0.094$\,mm$^{-1}$ (See Fig~\ref{pdfEk}). (e) Spatial spectrum of the energy $E(k)$ ($\bullet$). Contributions of gravitational $E_{grav} \sim S_h(k)$ and capillary $E_{cap} \sim k^2S_h(k)$ energies are well separated. The solid black line is the prediction for the capillary energy spectrum $\sim k^{-7/4}$. $\sigma_h = 3.6$\,mm and $\sigma_s=0.34$.}
    \label{sptcompWTb}
       \end{center}
 \end{figure} 

\section{Temporal and spatial spectra of wave elevation}
We next investigate the scaling of the spectrum with the spatial and temporal scales. Theoretically, the spectrum of capillary wave turbulence is predicted to scale as $S^{theo}_h (\omega) \sim \omega^{-17/6}$ and as $S^{theo}_h (k) \sim k^{-15/4}$~\cite{Zakharov1967,Pushkarev}.  Experimentally, the temporal spectrum of wave elevation $S_h (\omega) $ is obtained by averaging $S_h (\omega,k)$ over $k$ as shown in Fig.~\ref{sptcompWTb}(a) for different forcing amplitudes. When the forcing amplitude is increased, the peaks related to the forcing frequencies and their harmonics progressively disappear on the spectra, and a power-law spectrum is observed in the capillary regime on one decade in frequency ($20 \leq f \leq 200$\,Hz). No self-similar regime is observed in the gravity range due to the small size of the tank. Exponents from frequency power-law fits of spectra in the capillary range are plotted in Fig.~\ref{sptcompWTb}(b) as a function of the rms wave amplitude, $\sigma_h$. These spectral exponents are in a good agreement with the predicted exponent $f^{-17/6}$ at high enough wave amplitude. Note that the signal (recorded with the highest frame rate) reaches noise level around $200$\,Hz before reaching the dissipative range. These results are confirmed by simultaneous measurements provided by a capacitive wave gauge, which was used extensively in previous studies of capillary wave turbulence~\cite{Falcon2007,Falcon2010}. 

	The spatial spectrum $S_h (k)$ is obtained by averaging $S_h (\omega,k)$ over the frequencies $f$, and is shown in Fig.~\ref{sptcompWTb}(c) for different forcing amplitudes. For high enough wave amplitude, a power-law spectrum is also observed in the capillary range. Exponents from wave number power-law fits of spectra in the capillary range are plotted in Fig.~\ref{sptcompWTb}(d) and are close to the weak turbulence prediction in $k^{-15/4}$.  Finally, note that the departure from the theoretical power-law at small scales is observed in Fig.~\ref{sptcompWTb}(c) at ${\lambda_d}^{-1}$. ${\lambda_d}^{-1}$ is found to increase with an increasing forcing amplitude. This suggests that ${\lambda_d}^{-1}$ arises from the balance between viscous dissipation and nonlinear interactions~\cite{Zakharov1967,Deike}. Moreover, for the highest forcing amplitude, ${\lambda_d}^{-1}=0.48$ mm$^{-1}$ which corresponds to $f_d \sim 200$ Hz. For ${\lambda}^{-1}\geq {\lambda_d}^{-1}$, self similarity is broken and spectrum departs from the power-law. Although some strong hypotheses of weak turbulence (notably homogeneous, isotropic and infinite system) are not verified in our experimental system, correct agreement is found for both the temporal and spatial scalings of the capillary wave spectrum.

\begin{figure}[h!]
 \begin{center}
  \includegraphics[width=8.6cm]{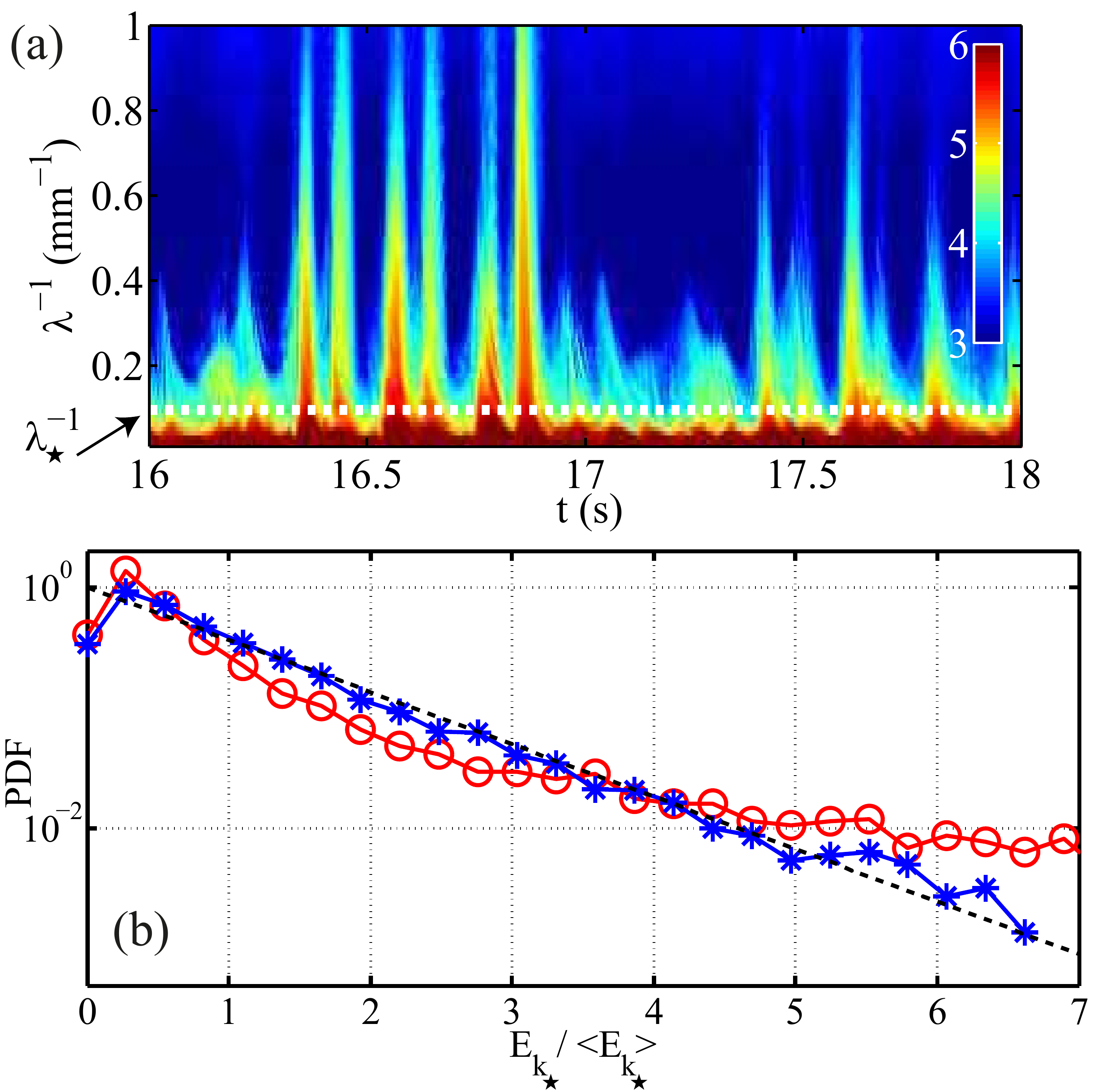}   
       \caption{(color online) (a) Evolution of the wave energy $E(k,t)$ in the ($t$,$\lambda^{-1}$) space. Color scale in $\log_{10}$ scale. (b) Probability density function of wave energy $E(k_{\star},t)/\langle E(k_{\star},t) \rangle$ at a mode $\lambda_{\star}^{-1}=0.094$\,mm$^{-1}$: ($\ast$) moderate forcing ($\sigma_h = 2.0\,$mm, $\sigma_s=0.19$), and ($\circ$) high forcing ($\sigma_h = 3.6\,$mm, $\sigma_s=0.34$). Dashed line is the exponential distribution: $\exp (-x)$.
}
    \label{pdfEk}
       \end{center}
 \end{figure}
 
\section{Statistics of Energy Modes}

We now use the energetic point of view to better probe the statistics of capillary wave turbulence. Assuming that, in average on a wave period, there is equality between the kinetic and potential energies at each scale, the total wave energy at one scale should thus be proportional to the potential energy at this scale. Therefore, neglecting non linear surface deformations, one can compute the total potential energy of the wave field as $$E_{pot}  \sim \int \left[ \rho g |\tilde{h}(k)|^2 +\gamma k^2 |\tilde{h}(k)|^2 \right]\mathrm{d}k \sim \int E(k) \mathrm{d}k\,.$$  From the spatial power spectrum $S_h (k) \sim |\tilde{h}(k)|^2$ ($\tilde{h}$ denotes the Fourier transform of $h$), one can thus easily deduce the energy spectrum $E(k)$ that is depicted in Fig.~\ref{sptcompWTb}(e). The separation of the respective contribution of the gravity energy and of the capillary energy appears clearly. Moreover, in the capillary range, $E(k)$ follows the predicted spectrum of weak turbulence for the energy $\sim k^{-7/4}$, confirming the above results. 

The spatial power spectrum $S_h(k)$ is now computed at each time step from the wave field $h(x,y,t)$. We thus obtain the temporal evolution of wave energy spectrum $E(k,t)$ as shown in Fig.~\ref{pdfEk}(a). An important result is that $E(k,t)$ exhibits stochastic bursts transferring energy towards small spatial scales. Indeed, most of the time, the wave energy is confined near the forcing scales, but from time to time, energy cascades through all spatial scales (within the inertial capillary range). Moreover, at a given spatial scale $k_{\star}$, i.e. a horizontal slice in Fig.~\ref{pdfEk}(a), $E(k_{\star},t)$ is found to be strongly erratic and bursts of random large-amplitude occur as for gravity wave turbulence \cite{Herbert2010}. The probability density function (PDF) of $E(k_{\star},t)$ of a mode $k_{\star}$ is shown in Fig.~\ref{pdfEk}(b) for two forcing amplitudes. For a moderate forcing, the PDF roughly follows an exponential distribution as expected for a Gaussian wave field~\cite{Nazarenkobook}. At higher forcing, a departure from the exponential distribution becomes significant, likely revealing a stochastic nonlinear phenomenon as described theoretically~\cite{Choi2005}. The PDF shape does not depend significantly on the scale $k_{\star}$ within the inertial range of the capillary cascade.

\section{Conclusion}
We have reported experiments on statistics of capillary wave turbulence generated by gravity waves. The full space and time resolved spectrum of wave height is obtained for the first time in the capillary range by means of an optical method. The spectrum exhibits both frequency and wave number power-law scalings whose exponents are close to the ones of the weak turbulence theory. Beyond the confirmation of these predictions in the capillary regime, significant departure from theoretical hypotheses has been reported (notably related to the isotropy and homogeneity of the wave field). We have also observed the occurrence of stochastic bursts in time transferring wave energy through the spatial scales within all the inertial range. The joint space and time resolved measurement have also allowed us to observe the linear and non linear dispersion relations of waves in a turbulent regime. When the forcing is strong enough, a nonlinear shift of the dispersion relation is reported, the spectrum power-law scalings with $k$ and $\omega$ being still in agreement with the predictions. Although the latter are derived using the linear dispersion, nonlinearity is considered in the theory as a perturbation of the linear state, and the observations reported here are thus consistent with the predictions. Further studies will investigate more in detail the statistics of Fourier modes of wave energy as well as the space-time correlations in order to better understand basic mechanisms involved in capillary wave turbulence.\\

\begin{acknowledgments} We thank N. Mordant for discussions and his help in data processing, M. Lesaffre for advices about optics of diffusing media, L. Deike for discussions and C. Falc{\'o}n for discussions. We acknowledge also A. Lantheaume and J. Servais for technical assistance. This work was funded by ANR-12-BS04-0005 Turbulon. 
\end{acknowledgments}


\end{document}